# EVOLUINDO RESILIÊNCIA EM ROTAS DE ÔNIBUS: PROPOSTA DE UM MÉTODO PARA A MAXIMIZAÇÃO DE ACESSIBILIDADE EM CENÁRIOS DE INCERTEZA POR MEIO DE ALGORITMO GENÉTICO

**André Borgato Morelli**
**André Luiz Cunha**
Universidade de São Paulo
Escola de Engenharia de São Carlos

**RESUMO**
Resiliência tem despertado interesse no campo do planejamento de transportes já que fenômenos raros, como crises de abastecimento de combustíveis, têm demonstrado potencial de desestabilizar sistemas de transporte. Contudo, os métodos propostos para planejar resiliência no transporte público deixam de levar em consideração o impacto que a frequência das rotas tem na acessibilidade do usuário. Para endereçar essa lacuna, este trabalho propõe um método de alocação de ônibus visando a maximização de acessibilidade em cenários de impacto – onde algumas rotas de ônibus têm sua frequência reduzida – a partir de um algoritmo genético. O método é aplicado no município de São Paulo e os resultados mostram que evoluir o sistema com a previsão de impactos moderados não apenas contribui para reduzir os efeitos negativos da redução da frequência de rotas, mas também melhora a sua eficiência em situação normal, mostrando a importância da contribuição desta pesquisa ao planejamento de sistemas eficientes.

**ABSTRACT**
Resilience has raised interest in transport planning as rare phenomena, such as fuel supply crises, have recently shown their potential to destabilize transport systems. However, the proposed methods for planning resilience in transit systems fail to consider the impact that bus frequency has on user accessibility. To address this gap, this paper proposes a bus allocation method aimed at maximizing accessibility in impact scenarios – where some bus routes have their frequency reduced – making use of a genetic algorithm. The method is applied in the city of São Paulo and the results show that evolving the system foreseeing moderate impacts not only contributes to reducing the negative effects of lower route frequency, but also improves its efficiency in normal conditions, showing the importance of the contribution of this research to the planning of efficient systems.

## 1. INTRODUÇÃO

A busca por resiliência em sistemas de transporte é um tópico que tem despertado interesse atualmente (Azolin *et al.*, 2020; Berche *et al.*, 2009; King *et al.*, 2020; Martins *et al.*, 2019; Mattsson e Jenelius, 2015). Isso se deve, em parte, à percepção de que eventos catastróficos, mesmo que raros, podem impactar de maneira significativa um sistema de transporte durante e após sua ocorrência. Um exemplo já explorado na literatura são crises de abastecimento de combustível (Azolin *et al.*, 2020; Martins *et al.*, 2019) que, apesar de terem curta duração, podem ter impactos duradouros na economia e nos sistemas de transporte. Outros eventos amplamente explorados são catástrofes naturais como inundações (Borowska-Stefańska *et al.*, 2019; He *et al.*, 2021) que não apenas têm certo caráter periódico, repetindo-se consistentemente em estações chuvosas, mas também tendem a se intensificar com a progressão da crise climática atual.

Nesse contexto, estudar resiliência em sistemas de transporte passa por identificar vulnerabilidades nos componentes desses sistemas (Furno *et al.*, 2018; Gecchele *et al.*, 2019; Jung *et al.*, 2020; Sarlas *et al.*, 2020). Em geral, esse processo se dá pela identificação de links mais importantes na rede, como no caso do trabalho de Berche *et al.* (2009) que propõe identificar seções críticas do sistema de transporte público a partir da centralidade de intermediação, medida derivada da teoria dos grafos.

Alguns trabalhos focaram especificamente no impacto de inundações nos custos de viagem enfrentados por usuários. Esse é o caso de Borowska-Stefańska *et al.* (2019) que estudaram o impacto de enchentes na acessibilidade da população de nove bacias hidrográficas na Polônia. Esse também é o caso no estudo de Kasmalkar *et al.* (2020) que mostraram como eventos de inundação que afetam a capacidade de vias podem impactar na eficiência de rotas de usuários e atrasos médios no sistema.

Outros trabalhos analisaram a substituição de meios de transporte em cenários de crise. Essa é a abordagem aplicada por Martins *et al.* (2019) ao estudar o potencial de substituição de viagens que consomem combustível fóssil por viagens por modos ativos através de distâncias máximas caminháveis/cicláveis. Azolin *et al.* (2020) incorporou o transporte público nessa análise ao identificar as linhas de transporte público que, ao serem desativadas, oferecem menor impacto ao sistema, economizando combustível no processo.

Enquanto o desenvolvimento de métodos para a análise de resiliência em transporte público tem se desenvolvido em ritmo acelerado, existem fraquezas marcantes nos métodos propostos até o momento. O primeiro caso, é o de análises voltadas à topologia da rede de transporte (Berche *et al.*, 2009; Shi *et al.*, 2019) que consideram "ataques" às redes de transporte público como remoções de links/nós (segmentos/estações de uma rota de trem ou ônibus). Esses são eventos que impactam primordialmente sistemas sobre trilhos, já que a danificação de um segmento impossibilita o trânsito em outros, mas é infrequente em sistemas de ônibus, a não ser nos que dependem de infraestrutura altamente especializada como BRTs. Desativar um ponto de ônibus ou até mesmo algumas vias sobre as quais os ônibus transitam pode causar algum impacto, mas, diferente de sistemas ferroviários, não impossibilitam sua operação.

Outro aspecto estudado é a desativação por completo de rotas de ônibus como no caso de Azolin *et al.* (2020). Enquanto essa é uma perspectiva mais realista do que acontece no sistema, o estudo dos autores não leva em consideração o fato de rotas mais longas necessitarem mais recursos para serem operadas (maior número de ônibus e gasto de combustível para manter uma frequência razoável), o que leva a uma supervalorização de rotas relativamente longas que atravessam mais zonas de tráfego, já que o maior acesso que essas rotas promovem não pe contrabalançado por seu maior custo e gasto de combustível.

Nesse caso, visando preencher a lacuna criada por estudos focados em resiliência de transporte público, este trabalho tem como objetivo avaliar a dinâmica existente entre a alocação de ônibus em uma linha e sua frequência. Isso possibilita a consideração de que, apesar de linhas mais longas de transporte público poderem atender um número elevado de pessoas, seu tempo de ciclo pode se alongar a ponto de requererem uma frota significativa para serem operadas em um nível de serviço razoável. Para tanto, propõe-se um algoritmo genético para evoluir uma solução que aloque a frota de ônibus disponível nas rotas existentes para amenizar eventos de redução de frequência ou desativação de linhas. Como medida de eficiência do sistema, adota-se a acessibilidade média a oportunidades de emprego, que depende não apenas da forma das linhas de ônibus, mas também dos *headways* dos veículos.

A maior inovação deste trabalho consiste no método proposto, que possibilita a tomadores de decisão alocar ônibus da frota de maneira racional e resiliente, possibilitando maior eficiência para um determinado número de veículos e, consequentemente, um dado nível de consumo de combustível. Além disso, este método pode ser aplicado para amenizar impactos de situações raras de abrupta redução de frota disponível para operação contínua, como crises de abastecimento de petróleo e paralisações por motivos variados.

## 2. MÉTODO PROPOSTO
### 2.1. Bancos de dados

A região explorada neste trabalho consiste nos bairros que compõem a Zona Oeste de São Paulo. As zonas de tráfego referentes a essa região foram filtradas da base original da pesquisa Origem/Destino de São Paulo de 2017 (Metrô-SP, 2017) com dados de empregos e população, como pode ser visualizado na Figura 1. A rede de transportes para pedestres foi obtida em forma de grafo orientado a partir da plataforma *OpenStreetMap* com auxílio da biblioteca OSMnx implementada em Python (Boeing, 2017). Já a rede de transporte público foi composta a partir do banco de dados da cidade em formato *General Transit Feed Specification* (GTFS) fornecido pela SPTrans (SPTrans, 2022) e fundida ao grafo de pedestres nos pontos de embarque (como mostrado na Figura 1).

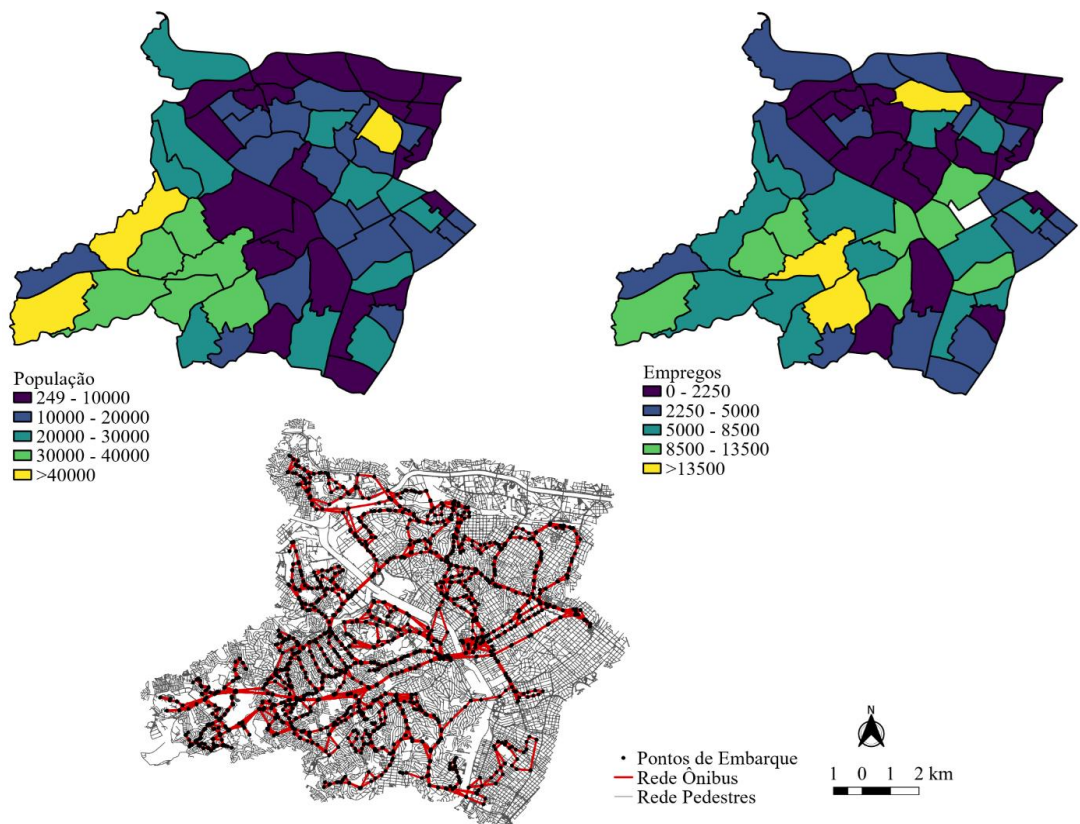

**Figura 1:** Características demográficas e econômicas (acima) e de rede (abaixo) da Zona Oeste de São Paulo.

Como este estudo é uma proposta inicial em uma área reduzida da cidade, são filtradas apenas as rotas de ônibus completamente contidas na Zona Oeste (que têm todos os pontos de embarque/desembarque na região), totalizando 45 rotas operando no período das 6h às 9h. Isso se deu para acelerar o treinamento dos sistemas para a avaliação de vários cenários diferentes e significa que grande parte da complexidade de interações entre as rotas dessa zona e outras mais longas ou pertencentes a outros modos como o metrô são perdidas. Na prática, isso significa que o modelo de transporte público utilizado neste trabalho não representa a condição real, mas sim uma situação sintética a partir da qual não se devem fazer julgamentos sobre o funcionamento do sistema de transporte público da Zona Oeste. Isso não implica problemas para este trabalho já que nossa proposta é a testagem de um método novo de otimização de sistemas de transporte público em um modelo que tenha as características básicas de um sistema de transporte público em tamanho reduzido. Ressalta-se que o ideal seria que o modelo fosse aplicado em um sistema real de cidade média, contudo os autores não tiveram acesso a uma base de dados GTFS do tipo no Brasil. Por fim, para constatar se o processo é válido para sistemas reais, ele foi testado no sistema todo da cidade de São Paulo, levando um tempo de processamento significativamente maior e, portanto, impossibilitando a avaliação de diferentes cenários.

**2.2. Análise de acessibilidade**

No século passado, acessibilidade foi um termo proposto por Hansen (1959) como "o potencial de oportunidades de interação". Desde então, diversas maneiras de medir acessibilidade surgiram, contudo elas podem ser categorizadas em três grupos principais: modelos acumulativos; modelos gravitacionais; e modelos utilitários (Geurs e van Wee, 2004). Métricas acumulativas são, em geral, as mais utilizadas na prática e em pesquisas (Carneiro *et al.*, 2019; Geurs e van Wee, 2004, 2004; Hanson e Schwab, 1987; Kwan, 1998; Kwan e Weber, 2003) e se destinam a contar o número de oportunidades que podem ser alcançadas a partir de um determinado ponto em um determinado tempo:

$$A_i = \sum O_j f(c_{ij}), \quad f(c_{ij}) = \begin{cases} 1 \; se \; c_{ij} \leq t \\ 0 \; se \; c_{ij} > t \end{cases} \quad \text{(Eq.1)}$$

Em que $A_i$ é a acessibilidade de $i$; $O_j$ é o número de oportunidades pertencentes à zona $j$, que neste trabalho é definido como o número de empregos da zona; $f(c_{ij})$ é a função de decaimento em relação ao custo de viagem ($c_{ij}$); e $t$ é o tempo limite de uma viagem.

Porém, neste trabalho foram consideradas métricas gravitacionais, que possuem formulação parecida com a das métricas acumulativas, mas podem ser mais bem ajustadas ao comportamento da população avaliada desde que exista uma pesquisa de desejo de viagem. Nesse caso, foi utilizada uma função de decaimento de formato log-logit modificada, já que se mostrou melhor ajuste na cidade de São Paulo quando comparada à gaussiana modificada e à exponencial inversa – outras funções tipicamente utilizadas nesses casos (Geurs e van Wee, 2004):

$$f(c_{ij}) = \frac{1}{1 + exp\left(-22.22 + 2.762 \, log\left(c_{ij}\right)\right)} \quad \text{(Eq.2)}$$

Com $c_{ij}$ sendo o tempo de viagem em segundos. O ajuste se deu por meio do método dos mínimos quadrados não linear sobre a curva de sobrevivência dos tempos de viagem. Para mais detalhes sobre o processo de ajuste, consultar Geurs e Ritsema van Eck (2001).

**2.3. Definição do problema para o algoritmo genético**

A proposta deste trabalho é otimizar um sistema de ônibus urbanos para situações de falha no suprimento do serviço, como quando a frequência das linhas é reduzida ou algumas linhas param de operar por motivos diversos, como paralisações ou crises de abastecimento de combustíveis. Para tanto, o primeiro passo foi definir o algoritmo de otimização e sua função objetivo.

Como pretende-se otimizar a acessibilidade a empregos, a função objetivo foi definida como a média da acessibilidade, ponderada pela população:

$$maximizar: \frac{\sum A_i P_i}{\sum P_i} \qquad (Eq.3)$$

Em que $P_i$ é a população residente na zona $i$ que não varia com a alocação de ônibus. Já a acessibilidade $A_i$ depende da eficiência da rede de transporte público que depende de sua forma, velocidade média, alcance e *headway*. Forma e alcance dependem da definição das rotas, aspecto que não será abordado neste trabalho, enquanto a velocidade média depende do número de paradas e condições gerais de tráfego que também não serão alteradas. O *headway* dos ônibus, por outro lado, depende diretamente de quantos veículos são alocados para operar em uma linha e de seu tempo de ciclo:

$$H = \frac{(TO + TT)}{O} \; segundos \qquad (Eq.4)$$

Em que $H$ é *headway* entre veículos; $TO$ é o tempo de trajeto do ônibus em segundos (inferido pelo GTFS); $TT$ é o tempo de terminal em segundos (adotado como uma média de 600 s neste trabalho); e $O$ é o número de ônibus operando a rota. Ao variar o número de ônibus em uma linha, portanto, é possível diminuir seu headway.

O *headway* impacta diretamente na capacidade de um usuário em acessar o sistema de transporte público. Neste trabalho, considera-se que um usuário tem que esperar em média $H/2$ segundos até que um veículo chegue e esse é o tempo de fricção associado à aresta do grafo conectando um nó do grafo de pedestres a um nó do grafo do transporte público (a aresta inversa, Transporte Público-Pedestre, possui custo zero).

Considerando que uma frota fixa opera na rede – calculada em 344 veículos a partir dos tempos de ciclo e *headways* definidos no GTFS de operação usual –, alocar mais ônibus em uma rota para reduzir seu *headway* significa reduzir o número de ônibus disponíveis para outras rotas, por isso a alocação racional de veículos, com as rotas mais importantes recebendo mais veículos, é importante para o bom funcionamento do sistema. Nesse caso, o problema que se tem em mãos é um problema de alocação de veículos nas rotas existentes de forma a maximizar a acessibilidade do sistema.

Para obter uma solução para esse problema, foi utilizado um algoritmo genético. Um código genético é definido como um vetor com comprimento igual ao tamanho da frota de ônibus (344 neste caso), com cada posição nesse vetor codificando a alocação de um ônibus em uma linha específica. Uma ilustração de um vetor genético que codifica a alocação de 10 ônibus em três rotas diferentes pode ser observada na Figura 2. O código genético possui quatro posições codificando para a rota A, cinco para a rota B e uma para a rota C, portanto as rotas A, B e C recebem quatro, cinco e um ônibus, respectivamente.

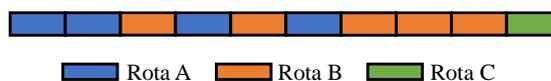

**Figura 2**: Vetor genético para um sistema de alocação. Cada cor representa uma linha. Quanto maior o número de genes codificando uma rota, maior o número de ônibus operando nessa.

### 2.3. Competição

O algoritmo genético funciona com a seleção de um grupo de soluções mais "aptas" para serem levadas à próxima geração. Existem diversas maneiras de conduzir uma competição (Goldberg e Deb, 1991; Lucasius e Kateman, 1994), mas neste trabalho foi utilizada a seleção por torneio (Goldberg e Deb, 1991), em que um evento seletivo compara três indivíduos aleatórios na população (pressão seletiva $k$ igual a 3), levando à geração posterior o indivíduo com maior função objetivo (Eq.4) entre os três. Isso significa que indivíduos com funções objetivo relativamente baixas têm alguma chance – mesmo que reduzida – de serem levados à próxima geração desde que estejam competindo em trios relativamente fracos, o que melhora a diversidade de indivíduos levados à geração posterior.

### 2.4. Mutação

Um evento importante no funcionamento do algoritmo genético é o processo de mutação (Hassanat *et al.*, 2019; Kumar *et al.*, 2010). Em cada geração, uma posição do código genético tem certa probabilidade (1% neste trabalho) de sofrer uma mutação. A mutação foi definida como uma alteração aleatória na rota à qual um ônibus é alocado, como ilustrado na Figura 3.

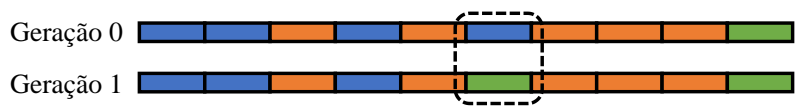

**Figura 3:** Ilustração de uma mutação em um vetor genético. Em um evento de mutação, uma posição do vetor recebe uma rota sorteada entre todas as possíveis.

### 2.5. *Crossover*

O processo de *crossover* é outro mimetismo biológico do algoritmo genético (Hassanat *et al.*, 2019). Nele, dois vetores que codificam uma alocação têm seu código genético compartilhado, de forma que a geração seguinte possua características de ambos. Neste trabalho, foi adotado o método de segmentação em ponto único (Hassanat *et al.*, 2019) que secciona os códigos genéticos em um ponto e recombina essas partes na próxima geração, como ilustrado na Figura 4. A probabilidade de *crossover* é fixada em 90% de forma que a maioria dos pares vencedores de competições tem seu código genético compartilhado, enquanto uma minoria é levada à próxima geração sem cruzamentos. Além disso, também foi utilizada a estratégia de elitismo em que o membro mais apto de cada geração é preservado sem alterações para a próxima geração, de forma a garantir que o indivíduo de melhor desempenho não se perca.

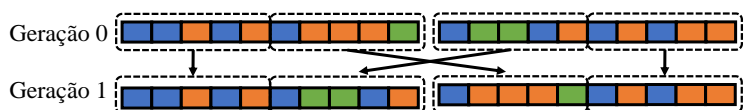

**Figura 4:** Ilustração do processo de crossover

### 2.6. Cenários avaliados

Ao aplicar o algoritmo genético, busca-se alocar os ônibus de forma a melhorar a acessibilidade do sistema e aumentar sua resiliência ao reduzir os impactos de eventos raros. Para tanto, é preciso incorporar processos de deterioração da rede de transporte público no modelo seletivo com intuito de forçar o sistema a evoluir de forma a mitigar esses fenômenos

Para tanto, foram avaliados três cenários distintos, um controle em que todas as linhas operam sem interrupções (cenário 1) e dois voltados a melhorar a resiliência do sistema. No cenário 2, uma falha minoritária no sistema foi simulada, em que cada linha tem 20% de chance de dobrar o seu *headway* em uma geração qualquer. Esse cenário simula problemas que podem acontecer em ônibus de uma determinada rota, como paralisações, crises de abastecimento de combustível ou manutenções não programadas que limitam a disponibilidade de veículos em operação. No cenário 3 uma falha majoritária no sistema foi simulada, em que cada linha possui 20% de chance de ser desativada completamente. Ressalta-se que as linhas que falham variam de geração a geração, de forma que o sistema deve evoluir resiliência sem saber a priori quais linhas falharão na próxima geração. Esse processo serve para forçar a evolução de um sistema apto a absorver qualquer evento negativo que possa acontecer na rede nessa proporção. Neste trabalho testou-se apenas a probabilidade de falha fixada em 20% dado o tempo de processamento requerido por cenário, mas em estudos futuros recomenda-se variar essa proporção.

## 3. RESULTADOS
### 3.1. Evolução dos modelos

A Figura 5 ilustra o processo de treinamento nos três cenários. A evolução se deu em 30 gerações com população igual a 48 seguida de 240 gerações com população igual a 12. Essa abordagem foi escolhida por garantir alguma diversidade no início da busca pelo ótimo, com maior número de redes, e uma evolução rápida no final. As populações foram selecionadas em múltiplos de 12 para serem compatíveis com o número máximo de processos suportados pela CPU utilizada. O tempo de execução do processo de cada cenário nas 270 gerações foi de 5,04h em média. Nota-se que no Cenário 1 o maior escore de acessibilidade em uma geração é estritamente crescente, o que acontece por conta do elitismo imposto ao modelo (o melhor de cada geração é sempre passado à frente). Esse escore inicia no patamar de 47.205 e termina no patamar de 59.461, apresentando um aumento relativo de acessibilidade de 26%. Nos outros cenários existe maior variação nos resultados, que acontece porque, apesar de o melhor integrante ser passado à frente, o impacto causado na rede é estocástico, com algumas gerações sofrendo impactos maiores que outras. Nota-se também que a variação no cenário 3 é mais abrupta já que os efeitos da desativação de rotas são mais significativos que os da redução da frequência de ônibus.

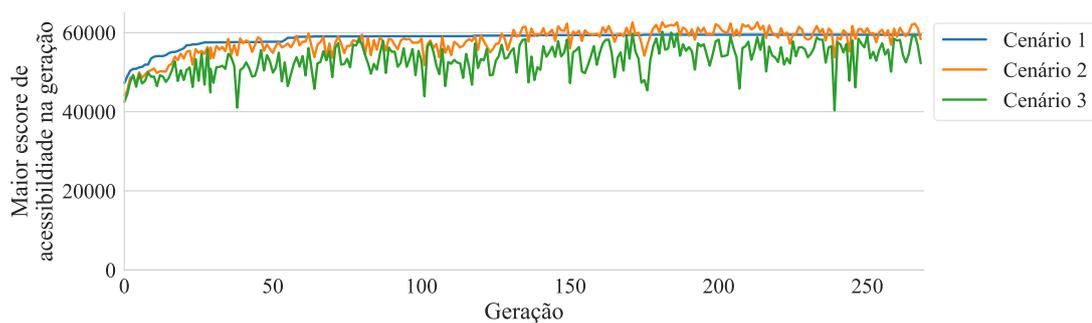

**Figura 5:** Evolução dos sistemas nos três cenários.

Essas linhas de tendência mostram o resultado durante o treinamento, portanto a linha do cenário 2 mostra a acessibilidade em uma rede com aproximadamente 20% das rotas com frequência reduzida e a linha do cenário 3 mostra a acessibilidade em uma rede com aproximadamente 20% das rotas desligadas. O resultado mais interessante está no fato de os cenários 2 e 3 superarem o escore do cenário 1 (que tem todas as rotas operando em máxima capacidade) em alguns momentos. Isso significa que, mesmo com algumas rotas sendo prejudicadas, esses sistemas ainda estavam operando com mais acessibilidade.

## 3.2. Teste em cenários de impacto

Para testar o desempenho dos melhores sistemas gerados pelos três cenários, foram simuladas 400 situações de impacto na rede de transporte público estudada. Em 200 dessas, foram simuladas reduções estocásticas de frequência (com chance de 20% de acontecer em uma linha qualquer). Nas outras 200, foram simuladas desativações estocásticas (com chance de 20% de acontecer em uma linha qualquer). Espera-se com isso testar o desempenho das redes treinadas para serem resilientes em relação à rede treinada no sistema estático.

A Figura 6 mostra as distribuições acumuladas e de densidade de probabilidade dos escores de acessibilidade para as situações simuladas. Nota-se que o sistema treinado no cenário 2 – impactos moderados – apresenta melhores resultados médios tanto quando a rede tem chance de 20% de ter o *headway* dobrado quanto quando a rede tem chance de 20% de ter rotas desativadas. É surpreendente que o sistema do cenário 3 seja superado pelo sistema treinado no cenário 2 mesmo quando avaliado em relação a desativação de rotas (cenário para o qual foi treinado).

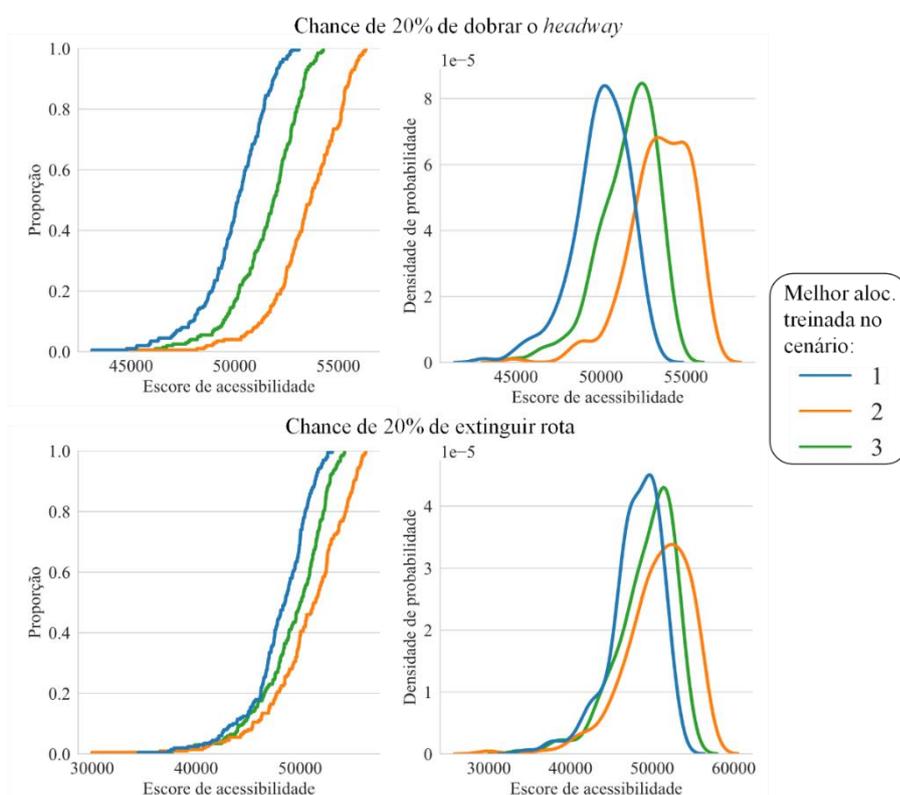

**Figura 6:** Distribuições acumuladas (esquerda) e de densidade de probabilidade (direita) para os três cenários.

A hipótese que se levanta é que a introdução de estresse moderado na rede faz com que ela fuja de mínimos locais e atinja patamares mais altos de acessibilidade, enquanto estresses maiores, apesar de terem o mesmo efeito, podem fazer com que o sistema fique muito errático na busca pelo ótimo, desacelerando sua convergência. Já o cenário sem estresse de rede mostrou estagnação rápida em um patamar mais baixo, o que sugere que o problema de alocação de ônibus para maximização de acessibilidade não seja convexo no ambiente de busca selecionado e possua mínimos locais.

Para testar a hipótese de pesquisa deste trabalho, de que é possível gerar sistemas de alocação de ônibus mais resilientes a partir de processos análogos à seleção natural, a Tabela 1 mostra os resultados do teste de Kolmogorov-Smirnov aplicado para testar a similaridade das distribuições dos resultados gerados no treinamento pelos cenários 1 e 2; e a diferença entre os cenários 1 e 3, demonstrando de maneira estatística a situação identificada nos gráficos da distribuição acumulada de frequência da Figura 6: ambos os cenários treinados para resiliência mostram uma diferença significativa na distribuição de seus escores em relação ao sistema treinado no cenário 1, com um deslocamento à direita (no sentido de aumento do escore médio em cenários de choque).

Além disso, a Tabela 2 contém os resultados do teste de Mood para a mediana dos dois cenários treinados para resiliência em relação ao cenário 1. O cenário 2 apresenta um aumento significativo de 5% a 6% na mediana em relação ao cenário um, enquanto o cenário três apresenta uma diferença menor, em torno de 3%, mas ainda significativa.

**Tabela 1**: Resultados do teste de Kolmogorov-Smirnov comparando os dois cenários resilientes (2 e 3) ao cenário controle (1)

| Falha (20% de chance) | Teste Kolmogorov-Smirnov | | |
|---|---|---|---|
| | Cenário Treino | Estat. | p-valor |
| **Aumento do** *headway* | 2 | 0,76 | 0,000 |
| | 3 | 0,44 | 0,000 |
| **Extinção** | 2 | 0,42 | 0,000 |
| | 3 | 0,27 | 0,000 |

**Tabela 2**: Testes para verificar significância de diferenças na mediana (Mood).

| Falha (20% de chance) | Teste Mood para a mediana das amostras | | | |
|---|---|---|---|---|
| | Cenário Treino | Dif. Relativa | Estat. | p-valor |
| **Aumento do** *headway* | 2 | -6,5% | 210,3 | 0,000 |
| | 3 | -3,3% | 56,3 | 0,000 |
| **Extinção da rota** | 2 | -5,6% | 37,2 | 0,000 |
| | 3 | -3,0% | 9,6 | 0,002 |

### 3.3. Resultados na rede completa de São Paulo

O gráfico da Figura 7 mostra a evolução do sistema de transporte público de São Paulo onde foi testado apenas o cenário 2 (de maior eficácia e eficiência). A rede consiste em 1.123 rotas de transporte público, 1.110 delas operadas por ônibus, e um total estimado em 14.653 ônibus na frota (a partir dos tempos de ciclo e *headways* das linhas). Esse é um exemplo de aumento drástico na complexidade do sistema, mas, apesar de o processo evolutivo desacelerar, ele continua efetivo, o que mostra que não são necessárias populações grandes para atingir bons resultados. Por outro lado, a execução das 270 gerações (30 com pop=48 e 240 com pop=12) levou um total de 42,6h no mesmo hardware utilizado anteriormente. Além disso, é possível que para que o sistema seja capaz de evoluir além do patamar mostrado na Figura 7, seja necessária uma população significativamente maior para conferir ao sistema a variedade genética necessária para evitar mínimos locais, o que aumentaria o tempo de processamento em computadores domésticos, talvez a ponto de inviabilizar a análise. Computadores comerciais, por outro lado, podem ser capazes de rodar o processo com diversidade em tempo factível.

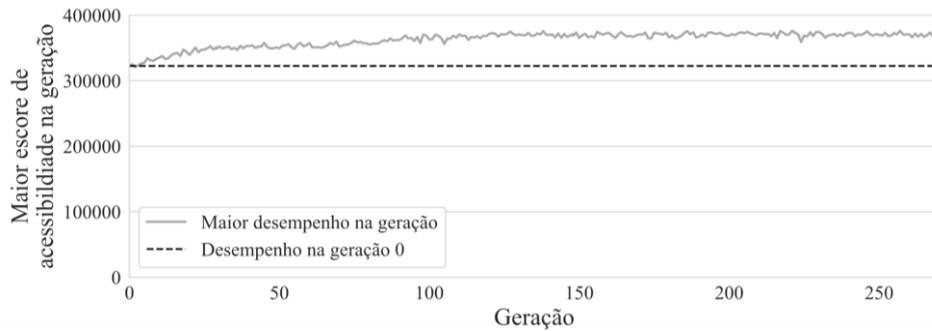
**Figura 7:** Evolução do sistema para a cidade de São Paulo completa (1110 linhas de ônibus, 14653 ônibus na frota).

## 4. CONCLUSÃO

Este trabalho propõe um método para aumentar a resiliência de sistemas de transporte público contra possíveis falhas por meio de otimização da alocação de ônibus em rotas existentes utilizando algoritmos genéticos. Para tanto, foi conduzido um estudo de caso de uma rede teórica compreendendo apenas as rotas de ônibus completamente contidas na Zona Oeste da cidade de São Paulo-SP, escolha que se deu pela dificuldade em se obter modelos GTFS de cidade menores no Brasil e a necessidade de uma região reduzida para a condução de teste iniciais do modelo. Posteriormente, o melhor cenário foi também testado na região completa de São Paulo.

Neste trabalho, objetivou-se otimizar a acessibilidade a empregos na região, ou seja, quanto maior o escore médio de acessibilidade de uma configuração do sistema, mais apta é considerada essa configuração. No método proposto, montam-se "códigos genéticos" para a operação do sistema de transporte público. Um código genético tem comprimento proporcional ao tamanho da frota de ônibus de forma que cada gene codifica a associação de uma unidade de ônibus a uma determinada rota. Quanto mais frequente o gene de uma rota em um código genético, maior o número de ônibus alocados nessa linha e, consequentemente, maior a frequência e qualidade do serviço. A partir dessa configuração, foi utilizado um algoritmo genético para evoluir um sistema mais apto em 270 gerações e três cenários diferentes. No cenário 1, todas as linhas de ônibus operam normalmente. No cenário 2, cada linha tem uma chance de 20% de sofrer uma falha minoritária em que a frequência dos ônibus alocados é reduzida pela metade. No cenário 3, cada linha tem uma chance de 20% de sofrer uma falha majoritária em que a linha deixa de operar.

Os resultados mostram que o algoritmo consegue aumentar significativamente a acessibilidade do sistema em todos os cenários testados. Um resultado surpreendente encontrado foi que pressionar o sistema de forma minoritária ou majoritária (cenários 2 e 3) faz com que o sistema não apenas evolua mais rapidamente e se torne mais resiliente a falhas, mas evita que o algoritmo fique estagnado em ótimos locais. Isso proporciona que o cenário evoluído com falhas estocásticas não apenas opere melhor em cenários de anômalos, mas que também opere melhor no cenário sem ocorrências. Como previsto, o sistema evoluído no cenário 2 se tornou mais capaz de absorver impactos minoritários, mas, diferente do esperado inicialmente, o sistema evoluído no cenário 3 não se tornou mais eficiente do que o sistema evoluído no cenário 2 em absorver impactos majoritários. A hipótese levantada para esse comportamento é de que uma pressão evolutiva minoritária consegue guiar o sistema lentamente a um ótimo global evitando ótimos locais, enquanto uma pressão evolutiva majoritária pode introduzir falhas grandes demais para favorecer o desenvolvimento acelerado do sistema. De toda forma, o cenário 3 ainda se mostra mais eficaz em evoluir o sistema a um ótimo resiliente do que o cenário 1 de controle.

A contribuição deste trabalho é avançar a literatura em técnicas para aumentar a eficiência e a resiliência de sistemas de transporte público pela ótica da acessibilidade. O método em seu estado atual beneficiaria principalmente cidades pequenas e médias já que o código não é otimizado para rodar de maneira veloz. Como exemplo, para o sistema considerado neste trabalho, com 344 ônibus em 45 linhas, foi possível rodar cada cenário de 270 gerações e população entre 12 e 48 em um período inferior a duas horas em um computador doméstico de seis núcleos de processamento. Nesse caso, é importante que sejam organizados esforços para catalogar os sistemas de transporte público nessas cidades brasileiras no formato GTFS, como já é comum em cidades europeias, e da América do Norte.

No caso da cidade completa de São Paulo, que conta com uma frota de quase 15 mil ônibus, são necessárias mais gerações e populações consideravelmente maiores para possibilitar a diversidade necessária para se alcançar um cenário otimizado, apesar de os resultados já mostrarem melhoras significativas para 270 gerações. O maior esforço computacional envolvido em avaliar cada membro da população em um sistema maior torna o processo bastante lento para computadores domésticos (~43h em nosso caso para alcançar uma solução subótima), mas computadores comerciais podem viabilizar essa aplicação.

Recomenda-se para estudos posteriores uma análise detalhada da alocação de ônibus otimizada para determinar quais rotas recebem maior atenção do sistema de otimização e por quais motivos. Além disso, é importante que sejam testadas diferentes funções de decaimento por distância ($f(c_{ij})$) para verificar seu impacto no estado final da alocação. Levanta-se a hipótese de que funções de decaimento mais abruptas (que valorizam tempos de deslocamento menores) tenderão a resultar em sistemas que valorizam aumentar significativamente a frequência em um conjunto reduzido de rotas, concentrando os ônibus alocados. Funções mais brandas como a função acumulativa, por outro lado, deverão proporcionar uma difusão maior dos ônibus nas rotas, com maior valorização da existência de rotas conectando diferentes pontos em vez da alta frequência dessas rotas. Ademais, os códigos desenvolvidos em linguagem Python de programação, bem como exemplos de aplicação, estão disponibilizados de maneira aberta[1].

---

[1] [link suprimido por conter informações sobre os autores].

André Borgato Morelli (andre.morelli@usp. br)
André Luiz Cunha (alcunha@ usp.br)
Departamento de Transportes, Escola de Engenharia de São Carlos, Universidade de São Paulo

Av. Trabalhador Sãocarlense, 400 – São Carlos, SP, Brasil